Kamil RZEPKA, Przemysław SKUROWSKI, Błażej ADAMCZYK
Silesian University of Technology, Institute of Informatics

Adam PILŚNIAK
Silesian University of Technology, Inst. of Measurement Science, Electronics and Control

# DESIGN OF PORTABLE POWER CONSUMPTION MEASURING SYSTEM FOR GREEN COMPUTING NEEDS

**Summary**. The article presents the design of a digital power measurement device intended for the green IT. Article comprises: use case analysis, accuracy and precision measurements and real life test of apache web server as exemplary application.

# PRZENOŚNY SYSTEM POMIARU ENERGII NA POTRZEBY ENERGOOSZCZĘDNYCH TECHNOLOGII INFORMACYJNYCH

**Streszczenie**. Artykuł omawia budowę cyfrowego miernika adresowanego na potrzeby 'Green IT'. Artykuł obejmuje: analizę przypadków użycia, pomiary dokładności i precyzji oraz przykładowe zastosowanie do analizy serwera www.

**Słowa kluczowe:** pomiar mocy i energii, zielone IT, zrównoważony rozwój

## 1. Introduction

Development of eco-friendly technologies and methods of saving the energy is one of the key challenges for scientists and engineers. Ecological technologies, that can reduce the emission of greenhouse gases is one of the research priorities of European Union [1].

Recent studies show that the reduction of costs becomes more important motivation for organizations [2]. As Green IT becomes more important for organizations, researches, that concerns reduction of the impact on the environment or even reduction of costs, are widely required. The payoff of efficient computer systems is twofold. Computers that consume less energy also throw off less heat. Therefore, they require less energy for cooling, which is really important in case of servers [3]. One of electricity suppliers in USA even offered rebates for buying efficient servers to reduce cooling loads, especially because of summer peak demand periods caused by air conditioning.

The intention of authors was to create a low cost, portable system of sufficient precision (class 1) which would be made of consumer grade parts. Moreover, open and extensible design makes the proposed solution easy to adopt for other applications. The article describes the analysis of requirements and design of a power measurement system comprising the digital registering device and a computer which acts as a controller. We present our solution in brief context of green IT needs. We identified two basic use cases to be handled in system design. Next we describe implementation details and measurement quality tests. Finally, as an operational verification, we used the system to measure power consumption of Apache server.

## 2. Green IT

A study and practice that includes designing, manufacturing together with using and disposing computer systems with minimal influence on the environment is called green IT [4]. Green IT can be considered from different points of view: engineering, institutional, metrology, manufacturing. These approaches are briefly described in following paragraphs.

First widely known initiative concerning the green IT was Energy Star labeling program. Its main goal was to promote energy-efficiency among consumer electronics. The TCO Certification program promoted CRT displays with low magnetic and electrical emissions.

In [5] the topic of measuring energy use of enterprise computer systems is covered. For this purpose a network of 266 custom built wireless sensors was deployed. The devices, for which the power consumption was measured, were divided into four classes: LCD screens, PCs, networking equipment, and servers. After two years of gathering the data and performing additional surveys the measurements were extrapolated to a whole building. The authors estimated that computing systems drew 50% of the building's energy load: 26% servers, 17% PCs, 4% displays, and 3.5% networking. Further analysis showed that even modification of a small details, like changing to dark desktop color schemes, can result in reduction of energy usage (in this case by 10%). In turn the network equipment consumed constant amount of power, even though it operated at various workload.

The authors of [6] investigated measurement methodology for possible mistakes that can occur during a measurement of power usage and analysis of the results. For this purpose they used a set of micro-benchmarks: FPU - floating point multiplication, INT - integer multiplication and division, Memory - wide-ranging random memory accesses, NOP - idle loop with NOPs, Cache - memory accesses with high cache-hit ratio. The study was conducted to confirm that the sampling rate and execution time, thermal effects, compilation configuration can have significant impact on the energy measurements. Another conclusion was that the memory events do not correlate with power consumption.

The work of Ardito and Morisio [7] focuses on summarizing and systematizing available knowledge on green IT. An Energy Life Cycle was introduced. It includes designing, manufacturing, transporting, using, disposing, and recycling of the IT product, because all of the activities consume materials and energy. Energy benchmarks and metrics were summarized into three categories: power, efficiency and productivity. Then energy consumption and carbon footprint of the global Information and Communication Technologies was assessed. ICT was supposed to be responsible for less than 5% of global electrical energy consumption, but the trend is increasing, due to the large increase in the number of individual IT devices. Finally a whole set of guidelines concerning energy efficient IT, including their pros and cons, was presented.

Another point of view on green IT comes from operational research. Węglarz in [1] describes models of scheduling problems additionally constrained by energy usage. There are defined two typical approaches which simplify analysis: laptop problem (fixed energy usage, optimizing scheduling), server problem (fixed scheduling, optimizing energy usage).

Energy-efficient design is crucial for battery-powered embedded systems. It demands optimizations both in software and hardware. In [8] the authors presented power manager module that cooperated with the applications and power-aware device drivers. It extended Red Hat ECOS, a real-time embedded operating system, running on an HP SmartBadge IV hardware platform. The implementation of application-driven power management policy resulted in energy reduction by more than 50% with no effect on performance. Another study performed on this hardware platform is described in [9]. Presented there source code optimizations, in some cases, offered up to 90% of reduction in energy consumption. On the other hand, the work [10] introduces domain-specific reconfigurable cryptographic processor. This energy-efficient algorithm-agile hardware provides the required flexibility, without incurring the high overhead costs associated with generic reprogrammable logic. The authors claimed that in ultralow-power mode the processor consumed at most 525 µW, which is comparable to dedicated hardware implementations, while providing all of the elasticity of a software-based implementation.

A study on the design of energy efficient microprocessors is presented in [11]. The authors defined metrics for energy efficiency for three modes of computation in digital circuits: fixed throughput, maximum throughput and burst throughput.

Other Green IT aspects are network and distributed systems related. Energy efficient network organization [12, 13] , transmission [14–16], routing [17] and cloud computing [18] are in focus of researchers. Energetic benchmarking of network services is also in scope of SPEC (Standard Performance Evaluation Corporation) which offers a SPECpower_ssj2008 [19] benchmark to evaluate the power and performance characteristics of server-side Java.

All of above require reliable energy consumption of computer systems measurements. The comprehensive review of five PC attached measurement devices is presented in [20]. These included both external and internal wattmeters. WattsUp was easy to use, out-of-box external meter and OmegaWatt was supposed to be mounted in a fuse box. The latter is an approach somewhat similar to ours but larger (6 channels), RS232 based and more expensive. Internal meters included: PowerMon2 which was designed to measure power usage of a computer's main board, NI9205 card in NIcDAQ-9178 chassis was used as a part of National Instruments general purpose Data Acquisition system, also a custom built DCM meter, based on a PIC microcontroller, was used to measure the instantaneous power at the 12 V ATX line. During the experiment a Linux specific benchmarks, that aimed at particular computer components (CPU, NICs, HDDs, memory), were used. The measurements were performed on desktop and server computers. Additionally, an analysis of an integrated power measurement interfaces was conducted.

## 3. Design of measurement environment

### 3.1. Use case analysis

The construction of the measurement environment requires two computers. All tests are carried on the *System Under Test* (SUT) thus its performance and energy usage is measured. The Controller's main responsibility is recording the power consumption. It can be also capable of gathering other results and sending workload requests to the SUT. The naming convention follows SPEC nomenclature [21].

During the design phase two use cases were identified. The first one occurs when a user launches a workload script on the Controller. Then the local power logging begins and the test is remotely started on SUT, i.e. via *ssh*. Fig. 1a shows this approach on an MSC diagram.

The second scenario envisages a user running a script workload directly on the SUT. Then the message to activate power logging is sent to the Controller. After the test finishes a second message, that terminates power logging, is sent. Respective MSC diagram is depicted in Fig. 1b

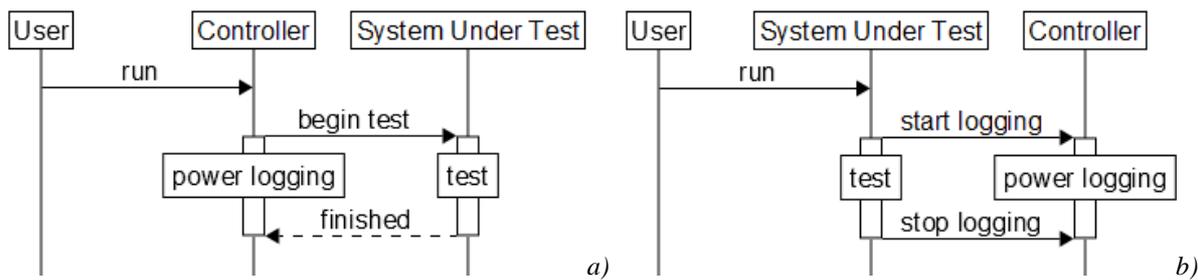

Fig. 1.  Use cases: a) remote test launching, b) remote power logging
Rys. 1.  Przypadki użycia: a) zdalne uruchamianie testu, b) zdalna rejestracja mocy

### 3.2. Design overview

The first idea of measuring the power consumption was to measure the DC current between a laptop and its power supply. The voltage is supposed to be constant, so the amount of energy could be obtained by multiplying it by the measured current and the time between the readings. Unfortunately the laptop was unable to boot while being connected through the meter. Measuring the DC current inside the case of a desktop computer is troublesome since it requires simultaneous data acquisition on many lines. Another drawback of this approach is not taken into account the power supply unit. Finally Orno OR-WE-504 was used as an external AC meter. The measuring device is mounted on the DIN rail with both RS485 ↔ USB converter and socket in small electric dedicated chassis (see Fig. 3c). This device is capable of measuring voltage, current, frequency, active, reactive and apparent power, power factor, active and reactive energy. The power consumption could be directly obtained from the measurement unit, although to ensure the calculation accuracy it was decided to employ generic formula for AC real power:

$$P = U \cdot I \cdot \cos\varphi, \tag{1}$$

$$E = P \cdot t, \tag{2}$$

where: $P$ is real power (Watts), $U$ is voltage (Volts), $I$ is current (Amps), $E$ is real energy (Joules), $t$ is time (seconds), and $\varphi$ represents the phase angle between the voltage and the current signals. The device offers the possibility to access the data through RS-485. Fig. 2a shows the general connection scheme of the measurement environment. Detailed scheme of the power meter is presented in Fig. 2b. Such connection allows for simultaneous voltage and current measurements. One of the design goals was to create an inexpensive, portable device. The case has 14 cm x 11 cm x 9 cm and weighs only 0.75 kg including 2.5 m power cable and 1.8 m USB cable. Fig. 2c shows the environment in use. Controller laptop is on the left and SUT computer is in the background. In the middle is the meter. Due to usage of RS-485 the OR-WE-504 can be also mounted in a fuse box (i.e. in a server room) and communicate with a computer through Ethernet class cable up to 1200 m.

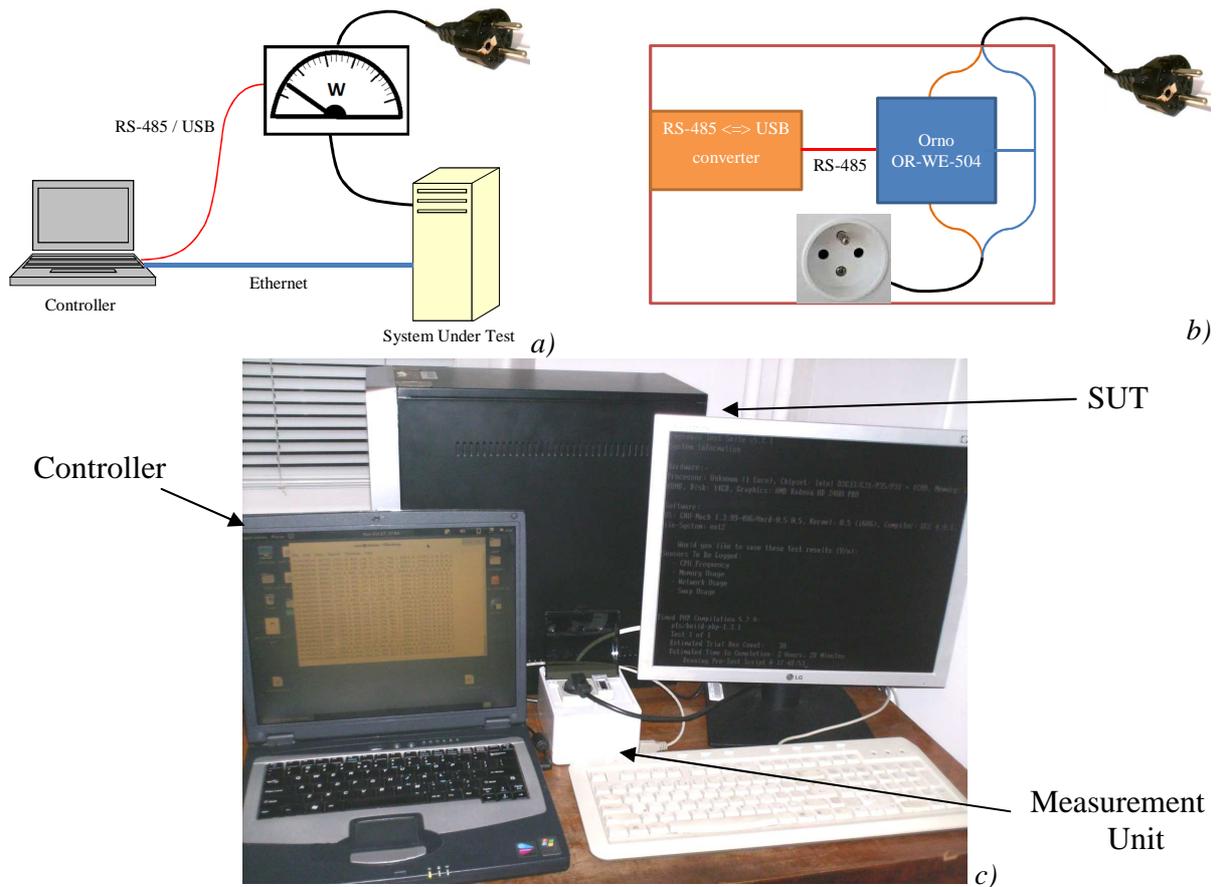

Fig. 2.   Illustrative connection scheme: a) external, b) internal, c) actual setup
Rys. 2.   Poglądowy schemat połączeń: a) zewnętrzny, b) wewnętrzny, c) rzeczywisty

### 3.3. Accuracy and precision of the meter

The manual of the power meter claims that the device has the accuracy class 1 (the error doesn't exceed 1%) [22]. It was decided to check whether it is really such accurate. For this purpose laboratory class (0.02) calibrator, Calmet C300, was used [23]. It gives the possibility to separately set the referential current and voltage, together with the phase shift between them. Table 1 shows obtained results. A legal energy meter is supposed to be checked on its base current (in case of OR-WE-504: 5 A), at one tenth of this value (0.5 A) and at four times this value (20 A). It was additionally checked how well the meter performs under inductive and capacitive load. What's more a legal energy meter should start indicating the power draw from 50 mA. OR-WE-504 started indicating the current from 95 mA so it doesn't meet the requirements of the standard. However, during measuring a common computer system, the power demand is unlikely to drop under 23 W. We can see that for an average value, in which preliminary power measurements were performed ($I = 237.4$ V, $U = 0.9$ A, $\cos\varphi = 0.792$), the device should be pretty accurate. Fig. 3 depicts the distribution of the measurements for given current and voltage. It can be noticed the power and energy errors are equal (Table 1). The time component of error doesn't influence much on results. It must be assumed that the accuracy class for measure of power is the same like for the energy meter - it means equal 1.

Table 1

Accuracy measurements

| No | Time [s] | $U$ [V] | $I$ [A] | Power factor [cos φ] | Provided energy [J] | Measured energy [J] | Relative energy error | Mean power [W] | Relative power error | Power variance |
|---|---|---|---|---|---|---|---|---|---|---|
| a) | 300.03 | 230 | 5 | 1 | 345032.93 | 343813.32 | -0.35% | 1146 | -0.35% | 0.0283 |
| b) | 170.01 | 230 | 0.5 | 1 | 19551.22 | 19502.76 | -0.25% | 114.7 | -0.25% | 0.0612 |
| c) | 94.02 | 230 | 20 | 1 | 432476.95 | 427962.15 | -1.0% | 4552 | -1.0% | 0.1286 |
| d) | 95.01 | 230 | 5 | 0.5 (ind.) | 54630.53 | 54792.51 | 0.30% | 576.7 | 0.30% | 0.0245 |
| e) | 95.01 | 230 | 5 | 0.5 (cap.) | 54629.41 | 53985.94 | -1.2% | 568.2 | -1.2% | 0.3237 |
| f) | 96.01 | 230 | 20 | 0.5 (cap.) | 220822.19 | 215548.10 | -2.4% | 2245 | -2.4% | 0.7722 |

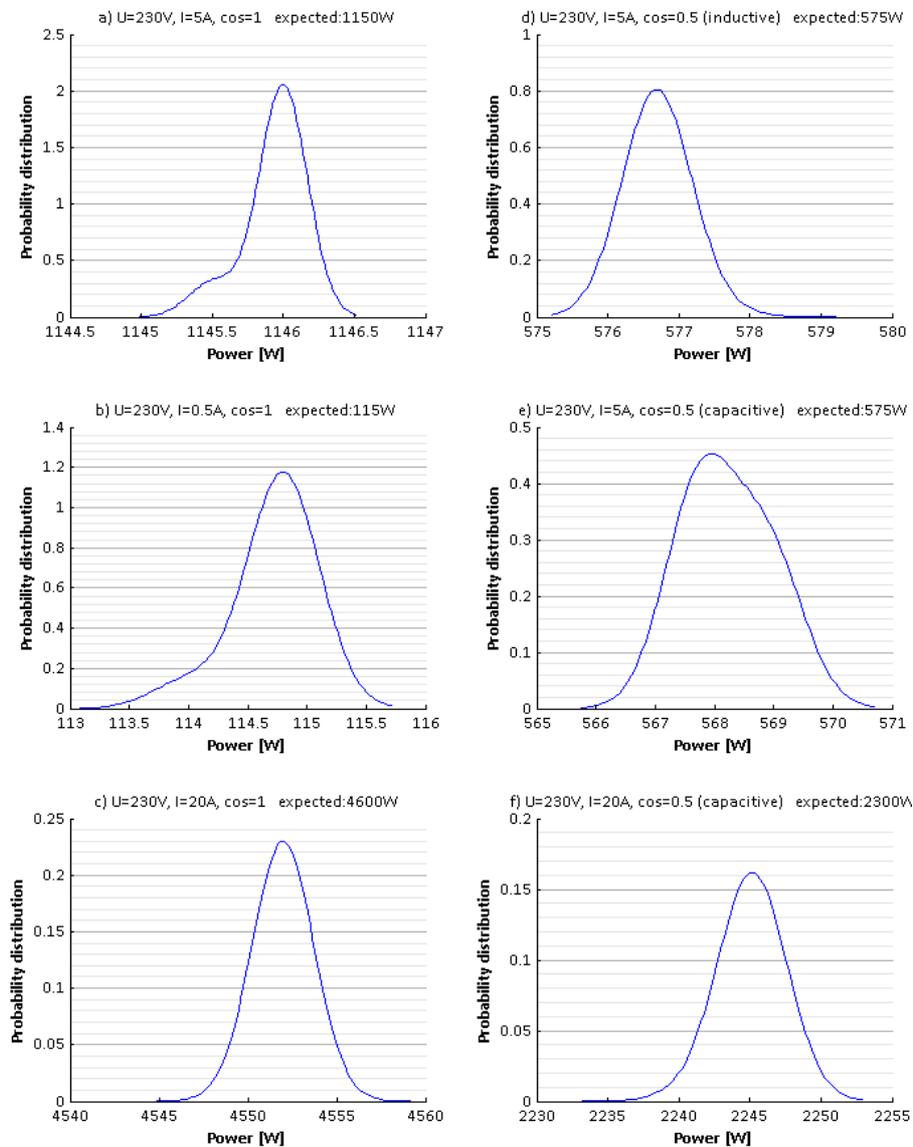

Fig. 3. Distribution of measurements of the reference energy consumption as in Table 1
Rys. 3. Rozkład pomiarów energii dla referencyjnego obciążenia jak w Tabeli 1

### 3.4. Software

The measurements were collected in Debian Linux using *Modpoll* program that implements the Modbus protocol [24]. It is a serial communications protocol and it is used in the industrial environment because of its simplicity and robustness. It was possible to tune all connection parameters including the frequency of readings from the program arguments. So it is a convenient tool to use from a command line or a script. Such usage had one drawback. Any redirection of *Modpoll*'s output stream resulted in its truncation. *Expect* tool was used to solve this problem [25]. It is used to automate interactions with interactive programs that expose a text terminal interface such as *telnet*, *ssh*. *Expect* uses pseudo terminals, starts the target program, and then communicates with it. This feature allowed for redirection of the *Expect*'s output stream, not the *Modpoll*'s itself.

As the *Modpoll* prints the measurements line by line (single measurement in line) it was required to develop a parsing program that changes the format to comma separated values. CSV is a standard format of the data, so the measurements can be imported to a spreadsheet for further analysis.

The communication protocol for remote logging, depicted in Fig. 2, was developed with the use of *netcat* and *xinetd*. It is simple stateful protocol with two system states – logging on and off, which are controlled by single TCP message containing `start|stop` commands that allowed to break the connection, just after sending the required messages, not waiting for the response. Such approach was necessary to avoid affecting the test results. Script, started on the SUT, was sending commands to the Controller using *netcat* – a simple utility that reads and writes data across network connections [26]. The Controller was running a server script on top of the *xinetd* super-server [27]. This daemon listens for requests from connecting clients and spawns a process which runs the appropriate executable. Table 2 presents the minimal necessary protocol which we developed. We didn't take the security issues into account as the system was designed to use in fully controlled laboratory environment. Any public usage would require implementing at least user authorization, which should be relatively simple.

Table 2
Minimal power logging control protocol

| command | unix command to issue |
|---|---|
| Begin power logging | `echo "start" | nc -q 0 controller_address controller_port` |
| End power logging | `echo "stop"  | nc -q 0 controller_address controller_port` |

In addition, an installation script responsible for enabling password-less communication between the Controller and the SUT was developed. Usage of RSA keys for authentication on remote server greatly simplifies running *ssh* in scripts.

## 4. Exemplary testing

To demonstrate the usefulness of the proposed solution we decided to perform a real life test of an apache web server. We used *JMeter* as a benchmark tool to generate arbitrary server workload. Both approaches - local and remote system testing were used to reveal the differences between two identified approaches. Since *JMeter* is CPU intensive application, because of Java utilization, so it allowed us to demonstrate the influence of the testing utility program on the results when it is launched on the tested hardware. On the other hand, workload stressing of the remote SUT would involve real network transmission, so it would follow real life usage. The tests followed the methodology provided by SPEC – first the maximum possible workload (100%) was identified. Having this value, intermediate workloads with 10% steps were generated. The last test was performed on the SUT with no workload – to measure the idle system power consumption. The results are presented in two ways. First, to demonstrate raw measured values – it is a time series of momentary power usage for a representative workload Fig. 4. Finally (Fig. 5), results are aggregated into averaged values for power consumption during the test and the system power productivity measured as requests per second per watt [req/s/W] which is equivalent to requests per joule.

The test Apache 2.2.22 webserver was deployed on a 64 bit GNU/Linux Debian 7.5 OS with kernel 3.2.0-4-amd64 and glibc 2.13. Test host was an average PC, comprising: • CPU - Intel Core 2 Quad Q6600 @ 3.0 GHz (4 cores) • Motherboard - Gigabyte P35-S3L, • RAM - 4 x 512MB @667MHz • HDD - 2x160GB Maxtor STM316021 • Power Supply - Mustang ATX KY-400W • VGA - HIS AMD Radeon 2400 PRO (working in text mode).

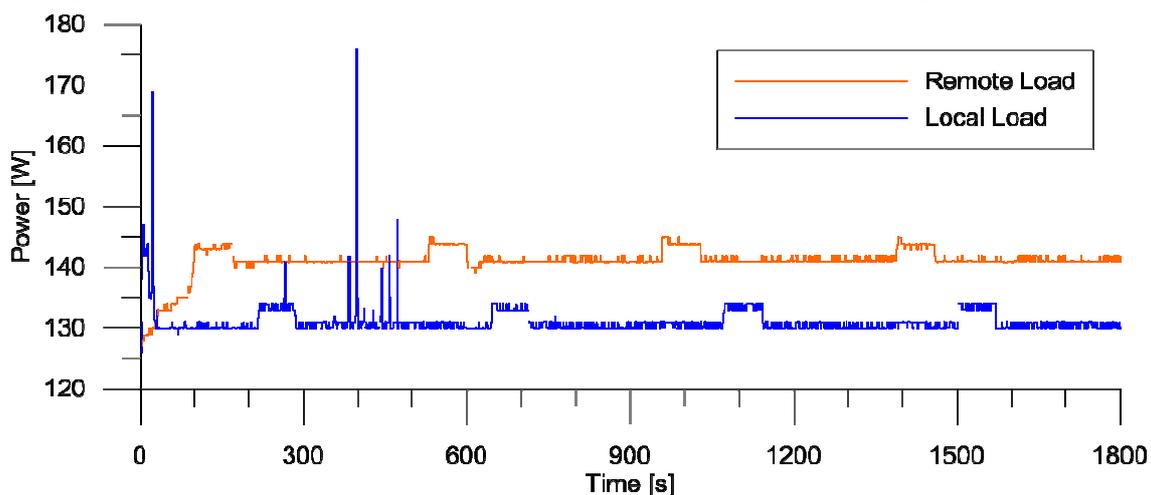

Fig. 4.   Time series of power consumption for 100% workload of apache for local and remote testing
Rys. 4.   Szereg czasowy poboru mocy  dla 100% obciążenia serwera przy lokalnym i zdalnym teście

At first glimpse (Fig. 4) local workload generator might seem to be more energy effective due to the smaller energy consumption (about 10 Watts), but considering only just power consumption is misleading. Launching of another, testing, process can affect overall server throughput so the maximal (100%) efficiency appeared to be different. To obtain real energetic efficiency it requires to take into account also absolute performance measure.

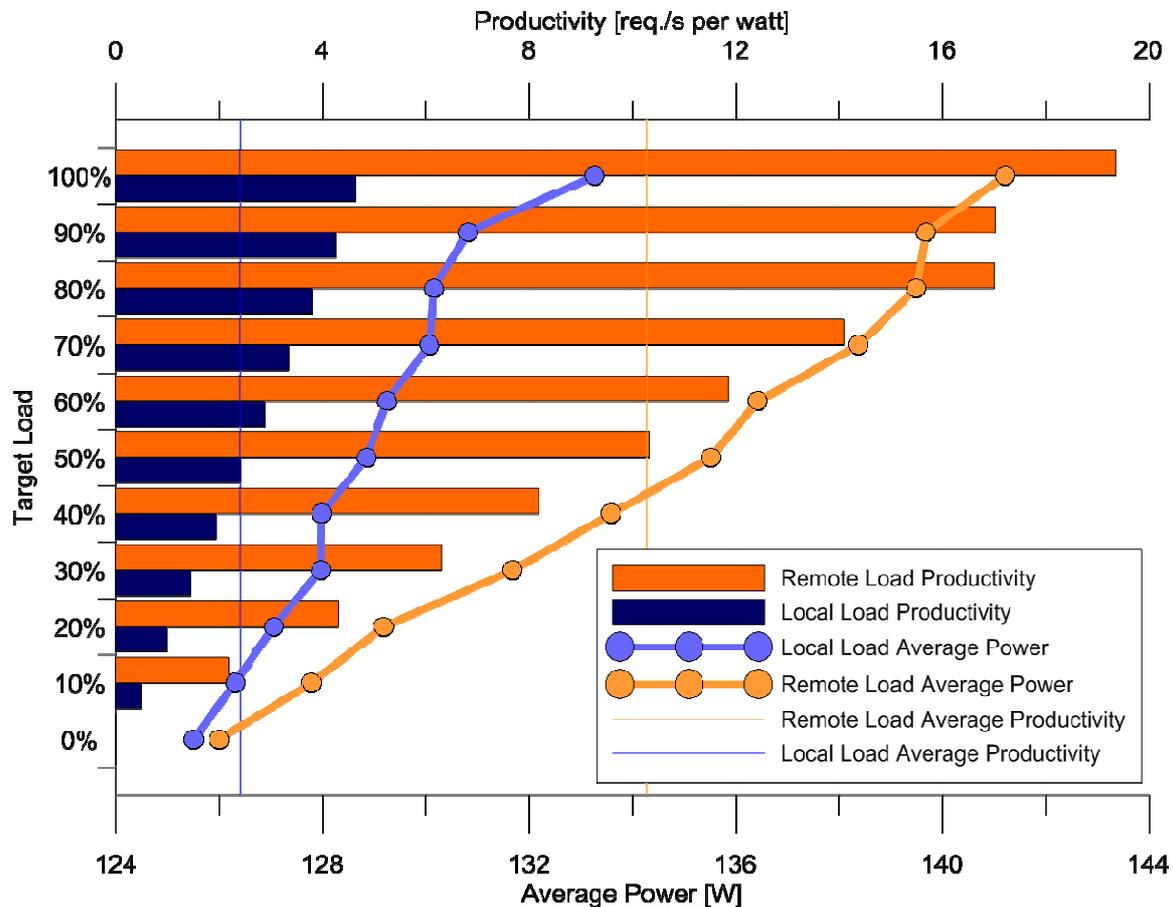

Fig. 5. Average power consumption and productivity for local and remote tests at 0-100% workload
Rys. 5. Średni pobór mocy i produktywność dla testu lokalnego i zdalnego przy obciążeniu 0-100%

The performed tests revealed the very significant influence of the measurement tool when it is started locally on the SUT. Despite loopback network connection between client and server, process switching and providing OS services to *JMeter* process reduced servers throughput capability. When compared to the remote test scenario the power demand of the SUT is reduced by certain factor but SUT productivity is affected far more ~4 times.

Obviously other, less CPU intensive workload generators, such as *ab* (apache benchmark) should affect the productivity less, nevertheless system testers should be aware about the fact and use appropriate testing method to evaluate the system. On the other hand there is plenty of local tests – measuring local computation performance (e.g. 7-zip LZMA test) – which must be started locally. Local testing with remote power logging should fit this kind of tests as the power overhead required to send a single UDP packet is negligible.

## 5. Summary

The provided system appears to be functional and low cost solution addressing the needs of a green IT. We demonstrated, with the exemplary experiment, the difference between (and the need to consider) the remote test launching and remote power logging scenarios. Another, notable feature is the simplicity of usage – to control power logging it requires to send single UDP packets which can be performed from the command line console. Finally in Table 3 we estimate the cost of the device which is another advantage of demonstrated solution.

Table 3
Approximate cost summary

| Part | approx. cost ( € ) |
|---|---|
| Measurement unit | 30 |
| Chassis (safety - IP30 compliant) | 3 |
| RS485 ↔ USB converter | 7 |
| Power socket + wiring + cord | 15 |
| TOTAL | 55 |

Further improvements might include: security extensions to enable logging in the public network, precise characterization of a measurement device for the measurement accuracy improvement by software bias cancellation. Possible further development, thanks to the organization of RS485, can include extending the device capabilities to register more than one power registering channel.

## Acknowledgements

This material is based upon work supported by the Polish National Science Centre under Grant No. N N516 479240. Authors also wish to thank dr Robert Wójcicki for his advices and sharing his experience in using power meter devices[1].

**BIBLIOGRAPHY**

1. Węglarz, J.: Wykład okolicznościowy o energooszczędnych technologiach informatycznych, in Profesor Jan Węglarz: doktor honoris causa Politechniki Śląskiej, Ziębiński, A. (ed.) Wydawnictwo Politechniki Śląskiej, Gliwice, 2014, pp. 21–29.

---

[1] http://www.mojazielonaenergia.pl/monitoring-on-line-instalacji-fotowoltaicznej/

2. Mines, C.: Green IT Adoption Is Driven by Business, Not Environmental, Considerations, *Chris Mines' Blog*. [Online]. Available: http://blogs.forrester.com/chris_mines/10-07-22-green_it_adoption_driven_business_not_environmental_considerations. [Accessed: 22-Sep-2014].
3. Du Bois, D.: What Is Green IT? Part 1: Cutting Emissions and Energy Use Enterprise-wide (Energy Priorities Archives), 04-Jun-2007. [Online]. Available: http://energypriorities.com/entries/2007/06/what_is_green_it_data_centers.php. [Accessed: 21-Oct-2014].
4. Green computing, Wikipedia, the free encyclopedia. 22-Sep-2014.
5. Kazandjieva, M., Heller, B., Gnawali, O., et al.: Measuring and analyzing the energy use of enterprise computing systems. Sustain. Comput. Inform. Syst., vol. 3, no. 3, Sep. 2013, pp. 218–229.
6. Mair, J., Eyers, D., Huang, Z., Zhang, H.: Myths in power estimation with Performance Monitoring Counters. Sustain. Comput. Inform. Syst., vol. 4, no. 2, Jun. 2014, p. 83–93.
7. Ardito, L., Morisio, M.: Green IT – Available data and guidelines for reducing energy consumption in IT systems. Sustain. Comput. Inform. Syst., vol. 4, no. 1, Mar. 2014, pp. 24–32.
8. Acquaviva, A., Simunic, T., Benini, L.: LP-ECOS: An Energy Efficient RTOS. Hewlett-Packard, Technical Report HPL-2003-81, Apr. 2003.
9. Simunic, T., Benini, L., De Micheli, G.: Energy-efficient design of battery-powered embedded systems. IEEE Trans. Very Large Scale Integr. VLSI Syst., vol. 9, no. 1, Feb. 2001, pp. 15–28.
10. Goodman, J., Chandrakasan, A.P.: An energy-efficient reconfigurable public-key cryptography processor. IEEE J. Solid-State Circuits, vol. 36, no. 11, Nov. 2001, pp. 1808–1820.
11. Burd, T.D., Brodersen, R.W.: Energy efficient CMOS microprocessor design. in Proceedings of the Twenty-Eighth Hawaii International Conference on System Sciences, 1995, 1995, vol. 1, pp. 288–297 vol.1.
12. Kim, K.-H., Min, A.W., Gupta, D., et al.: Improving energy efficiency of Wi-Fi sensing on smartphones. in 2011 Proceedings IEEE INFOCOM, 2011, pp. 2930–2938.
13. Blume, O., Zeller, D., Barth, U.: Approaches to energy efficient wireless access networks. in 2010 4th International Symposium on Communications, Control and Signal Processing (ISCCSP), 2010, pp. 1–5.

14. Bathula, B.G., Elmirghani, J.M.H.: Green networks: Energy efficient design for optical networks. in IFIP International Conference on Wireless and Optical Communications Networks, 2009. WOCN '09, 2009, pp. 1–5.
15. Gomez, K., Riggio, R., Rasheed, T., Granelli, F.: Analysing the energy consumption behaviour of WiFi networks. in 2011 IEEE Online Conference on Green Communications (GreenCom), 2011, pp. 98–104.
16. Vergara, E.J., Nadjm-Tehrani, S., Prihodko, M.: EnergyBox: Disclosing the wireless transmission energy cost for mobile devices. Sustain. Comput. Inform. Syst., vol. 4, no. 2, Jun. 2014, pp. 118–135.
17. Andrews, M., Anta, A.F., Zhang, L., Zhao, W.: Routing and Scheduling for Energy and Delay Minimization in the Powerdown Model. in 2010 Proceedings IEEE INFOCOM, 2010, pp. 1–5.
18. Lawey, A.Q., El-Gorashi, T., Elmirghani, J.M.H.: Energy efficient cloud content delivery in core networks. in 2013 IEEE Globecom Workshops (GC Wkshps), 2013, pp. 420–426.
19. SPEC power_ssj2008. [Online]. Available: http://www.spec.org/power_ssj2008/. [Accessed: 23-Sep-2014].
20. Diouri, M.E.M., Dolz, M.F., Glück, O., et al.: Assessing Power Monitoring Approaches for Energy and Power Analysis of Computers. Sustain. Comput. Inform. Syst., vol. 4, no. 2, Jun. 2014, pp. 68–82.
21. -: SPEC Power and Performance. Benchmark Methodology V2.1. Standard Performance Evaluation Corporation.
22. OR-WE-504_PL - OR-WE-504_PL_manual.pdf. [Online]. Available: http://www.orno.pl/grafiki2/INSTRUKCJE/WSKAZNIKI%20ENERGII/OR-WE-504_PL_manual.pdf. [Accessed: 15-Oct-2014].
23. Three Phase Power Calibrator and Device Tester - C300-data-sheet-EN.pdf. [Online]. Available: http://www.calmet.com.pl/pdf/C300-data-sheet-EN.pdf. [Accessed: 16-Oct-2014].
24. Modbus, Wikipedia, the free encyclopedia. 20-Sep-2014.
25. Expect, Wikipedia, the free encyclopedia. 24-Aug-2014.
26. Netcat: the TCP/IP swiss army. [Online]. Available: http://nc110.sourceforge.net. [Accessed: 04-Nov-2014].
27. xinetd, Wikipedia, the free encyclopedia. 01-Nov-2014.

**Omówienie**

Artykuł omawia przenośny system do pomiaru zużycia energii elektrycznej (AC 230 V) w systemach komputerowych. System powstał na potrzeby oceny efektywności energetycznej (przyjazności dla środowiska) i zbudowany jest w oparciu ogólnodostępne (konsumenckie) i niedrogie części. Analiza scenariuszy użycia wykazała dwa podstawowe schematy (Rys. 1): testowanie zdalnego systemu, gdzie kontroler poprzez sieć generuje obciążenie i mierzy potrzebną moc w testowanym systemie; testowanie lokalne ze zdalną rejestracją mocy, gdzie testowany system powiadamia kontroler o rozpoczęciu i zakończeniu eksperymentu.

Zaproponowane rozwiązanie składa się z przyrządu pomiarowego połączonego złączem USB do przenośnego komputera i oprogramowania umożliwiającego akwizycję danych pomiarowych w obydwóch powyższych scenariuszach, przy czym protokół komunikacyjny jest trywialnie prosty i możliwy do zaimplementowania poziomu konsoli (patrz Tabela 2). System pomiarowy nie wymaga ingerencji w badany system - potrzeba jedynie wpięcia zasilania poprzez urządzenie pomiarowego pomiarowe (Rys. 2).

Artykuł omawia przeprowadzone testy dokładności i precyzji pomiaru urządzenia pomiarowego (Tabela 1, Rys. 3) przeprowadzone w specjalistycznym laboratorium wykorzystując obciążenie wzorcowe (Calmet) o 50-krotnie większej dokładności.

Finalnie w artykule zademonstrowano użycie proponowanego systemu do oceny wydajności energetycznej (w liczbie żądań na dżul) serwera WWW Apache. W trakcie eksperymentu zademonstrowano różnice w pomiarach zużycia energii i wydajności energetycznej pomiędzy wywołaniem lokalnym i zdalnym testem. Wykazano celowość zdalnego generowania obciążenia dla usług sieciowych. W podsumowaniu zestawiono koszty (Tabela 3) zbudowania urządzenia pomiarowego, wskazano również możliwości dalszego rozwoju: programowej poprawy dokładności pomiarów, rozbudowania systemu o mechanizmy bezpieczeństwa, oraz wielokanałowego systemu rejestracji poboru mocy.

**Addresses**

Kamil RZEPKA: Silesian University of Technology, Institute of Informatics, ul. Akademicka 16, 44-100 Gliwice, Poland, kamil.rzepka@op.pl

Przemysław SKUROWSKI: Silesian University of Technology, Institute of Informatics, ul. Akademicka 16, 44-100 Gliwice, Poland, przemyslaw.skurowski@polsl.pl.

Błażej ADAMCZYK: Silesian University of Technology, Institute of Informatics, ul. Akademicka 16, 44-100 Gliwice, Poland, blazej.adamczyk@polsl.pl .

Adam PILŚNIAK: Silesian University of Technology, Inst. of Measurement Science, Electronics and Control, ul. Akademicka 19, 44-100 Gliwice, Poland, adam.pilsniak@polsl.pl